# HARMONIZATION MITIGATES DIFFUSION MRI SCANNER EFFECTS IN INFANCY: INSIGHTS FROM THE HEALTHY BRAIN AND CHILD DEVELOPMENT (HBCD) STUDY

*Elyssa M. McMaster[1], Gaurav Rudravaram[1], Michael E. Kim[1], Trent M. Schwartz[1], Chloe Scholten[2], Jongyeon Yoon[1], Adam M. Saunders[1], Andre T. S. Hucke[1], Karthik Ramadass[1], Emily M. Harriott[1], Steven L. Meisler[3], Simon N. Vandekar[4], Allen Newton[1,4], Seth A. Smith[1,4], Saikat Sengupta[4], Kathryn L. Humphreys[1], Sarah Osmundson[4], Daniel Moyer[1], Laurie E. Cutting[1,4], and Bennett A. Landman[1,4]*

[1]Vanderbilt University, Nashville, TN, USA, [2]University of Calgary, Calgary, AB, Canada, [3]University of Pennsylvania, Philadelphia, PA, USA, [4]Vanderbilt University Medical Center, Nashville, TN, US

## ABSTRACT

The HEALthy Brain and Childhood Development (HBCD) Study is an ongoing longitudinal initiative to understand population-level brain maturation; however, large-scale studies must overcome site-related variance and preserve biologically relevant signal. In addition to diffusion-weighted magnetic resonance imaging images, the HBCD dataset offers analysis-ready derivatives for scientists to conduct their analysis, including scalar diffusion tensor (DTI) metrics in a predetermined set of bundles. The purpose of this study is to characterize HBCD-specific site effects in diffusion MRI data, which have not been systematically reported. In this work, we investigate the sensitivity of HBCD bundle metrics to scanner model-related variance and address these variations with ComBat-GAM harmonization within the current HBCD data release 1.1 across six scanner models. Following ComBat-GAM, we observe zero statistically significant differences between the distributions from any scanner model following FDR correction and reduce Cohen's *f* effect sizes across all metrics. Our work underscores the importance of rigorous harmonization efforts in large-scale studies, and we encourage future investigations of HBCD data to control for these effects.

## 1. INTRODUCTION

Large-scale national studies introduce an exciting opportunity to study population-level brain qualities and changes [1], [2], but these studies inherently introduce site-related biases despite best efforts for protocol harmonization due to scanner hardware limitations [3], [4], [5]. The HEALthy Brain and Childhood Development (HBCD) Study [1], [2] is an ongoing multi-site, longitudinal initiative to collect imaging data as well as other data of interest (biospecimens, surveys, etc.) at unprecedented scale for infancy and early childhood. Since the HBCD consortium spans 27 recruitment sites [1] across the United States, scientists must anticipate significant site and scanner-related variance.

HBCD's diffusion MRI protocols include high resolution spatial sampling, uniform TR and TE, and a common multi-shell diffusion acquisition, however, some parameters, such as flip angle and phase partial Fourier vary across scanners [1]. Though the pre-harmonized parameters from HBCD show great promise for future white matter modeling [6], [7] the diffusion-relevant scanner effects for this study have not yet been investigated despite the use of six unique scanner models.

All data for this experiment is accessible in release 1.1 of HBCD. Participants cluster around the same age at time of scan, which skews the current available data to include 437 subjects with available HBCD site codes younger than 0.2 years old, each scanned on one of six scanner models (Fig. 1). HBCD's data includes derivates processed with *QSIPrep* and *QSIRecon* [8] including calculation of diffusion tensor imaging (DTI) scalar metrics within bundles derived from TORTOISE tensor fitting [9], [10]. These DTI metrics include axial diffusivity (AD), fractional anisotropy (FA), lattice index (LI), and radial diffusivity (RD) [9]; population-level analysis of these metrics inform biomarkers of disease and disorder [11] and harmonization within infants offers an opportunity to characterize expected ranges for a healthy population in a pediatric cohort.

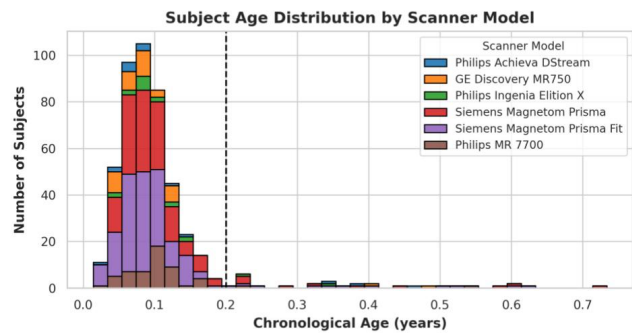

**Fig. 1.** The data available from HBCD Release 1.1 clusters between 0 and 0.2 years old. We exclude subjects over 0.2 years of age to avoid additional variation.

In this work, we quantify the scanner model-related effects within the distribution of DTI metrics. Despite the presence of scanner biases in DTI metrics, we hypothesize that ComBatGAM harmonization [4], [5] mitigates unwanted site-related variation in the HBCD derived data.

## 2. METHODS

HBCD collected diffusion-weighted images (DWI) from the 437 subjects of interest across six scanner types with spatial resolution 1.7 mm$^3$, TR 4800 ms, TE 88 ms, and $b$ values of 0, 500, 1,000, 2,000, and 3,000 s/mm$^2$. Data underwent preprocessing and multi-shell tensor modeling via TORTOISE [10] to generate scalar maps [9].

Following collection of HBCD scalar metrics from the LASSO data sharing platform, we applied ComBat-GAM as presented by Pomponio et al. [5] based on the masked mean for four scalar metrics, axial diffusivity (AD), fractional anisotropy (FA), lattice index (LI), and radial diffusivity (RD) in every bundle for the 437 subjects that fit our study criteria (available site label and between 0 and 0.2 years old, inclusive). We select ComBat-GAM over traditional ComBat to handle biologically related variability within the scalar metrics expected with brain development [4], [5], [11].

ComBat-GAM assumes that the input scalar features for feature $f$ from scan $i$ from scanner model $s$ are defined in Equation (1) as

$$Y_{isf} = \alpha_f + f_f(age_i) + \beta_{f,sex}Sex_i + \gamma_{sf} + \delta_{sf}\varepsilon_{isf} \quad (1)$$

where $\alpha_f$ is the global mean for feature $f$ across all subjects and scanner models, $f_f(age_i)$ is the smooth nonlinear age function to capture how feature $f$ changes with age, fitted with a smooth spline to preserve developmental trends across sites, $\beta_{f,sex}$ is the linear covariate to capture the average male/female difference in feature $f$, interpreted as the mean offset between the sexes following age adjustment, $Sex_i$ is the predictor (coded 0 for male and 1 for female), $\gamma_{sf}$ is the site additive site effect, or site-specific bias, for the mean of feature $f$, $\delta_{sf}$ is the scaling term to model the variability of $Y$ caused by site $s$ for feature $f$, and $\varepsilon_{isf}$ is an error term [12].

Following ComBat-GAM harmonization, we use ANOVA statistical tests to evaluate statistically significant differences between each scanner model's distribution for the four metrics and the 67 bundles defined by HBCD [1] and follow the tests with false discovery rate (FDR) correction per metric. We compute the absolute value of Cohen's $f$ effect size [13]. Cohen's $f$ effect size defines a small effect as $f = 0.10$, a medium effect as $f = 0.25$, and a large effect as $f = 0.40$; for this experiment, we define small effect size as $f \leq 0.1$, medium effect size as $0.1 < f \leq 0.25$, and large effect size as $f > 0.25$.

## 3. RESULTS

The ANOVA test resulted in 221 positive ($p < 0.05$) results out of 268 tests (4 metrics x 67 bundles) prior to ComBat-GAM harmonization after FDR correction, which indicates that at least one scanner model had statistically significant effects within most bundles for most metrics. We show the $\gamma_{sf}$ (shifting factor) and $\delta_{sf}$ (scaling factor) for every scanner model and metric combination in Figure 2.

We perform the same statistical tests following ComBat-GAM harmonization and observe zero statistically significant scanner model effects following FDR correction. We evaluate mean Cohen's $f$ effect size for each bundle across all scanner models. Before ComBat-GAM harmonization, the Cohen's $f$ effect size computations result in 13 (4.9%) results with small effect size, 147 (54.9%) results with medium effect size, and 108 (40.3%) results with large effect size. Following ComBat-GAM, the Cohen's $f$ effect size computations result in 266 (99.3%) results with small effect size, 2 (0.7%) results with medium effect size, and 0 results with large effect size. (Fig. 3). Spatial maps from HCP-842 [14] with bundles registered to MNI-152 [15] reveal that different bundles experience different scanner-related effect sizes across the brain for FA, but ComBat-GAM harmonization reduces these effects across the brain (Fig. 4).

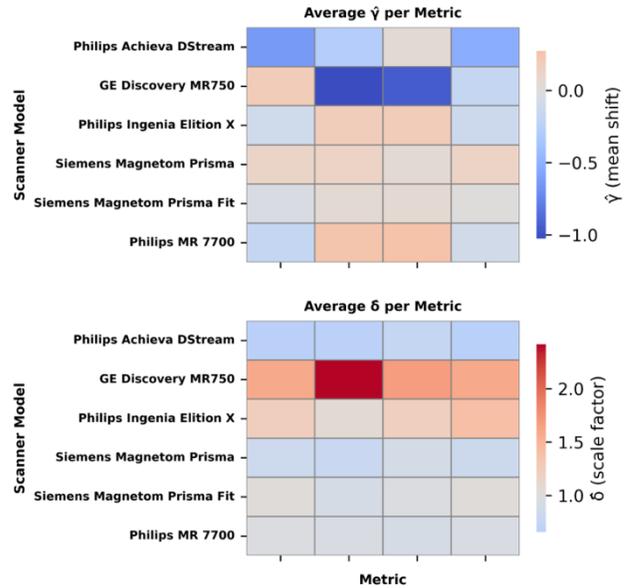

**Fig. 2.** Average gamma and delta coefficients for all bundles for every scanner model and metric combination following ComBat-GAM. Notably, the bias correction needed for every individual scanner type varied by metric, with no one scanner type consistently inducing the most variance.

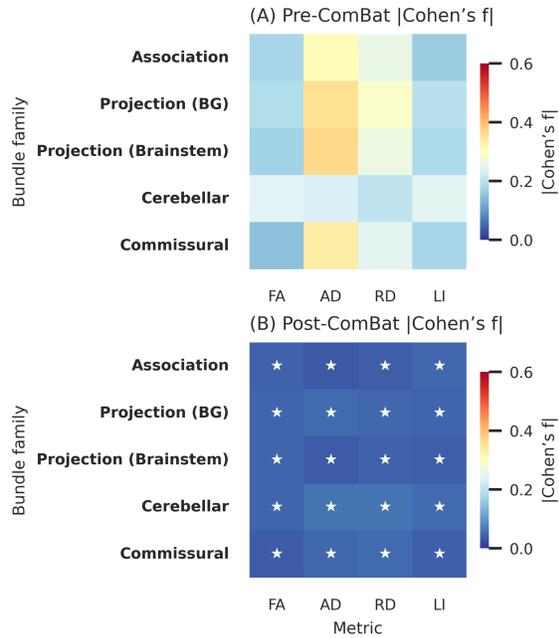

**Fig. 3.** Observed median Cohen's *f* effect size for all four metrics across the five HBCD-defined major bundle before and after ComBat-GAM (small effect sizes denoted with a star; successful site variation reduction is indicated with deeper blue tones). Median effect size minimizes across the brain, suggesting effective cross-scanner harmonization.

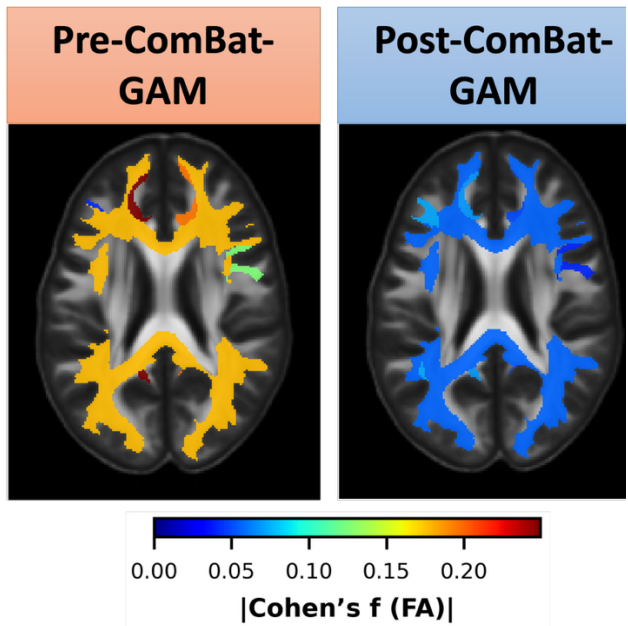

**Fig. 4.** A representation of FA effect size before and after ComBat-GAM for 5 bundles (Association_ArcuateFasciculusL/R, Association_CingulumL/R, and the Commissure_CorpusCallosum) overlaid on MNI-152 for visibility. Though bundle effect sizes vary from medium to large across the brain prior to harmonization, the post-ComBat-GAM spatial map shows small effect size.

## 4. DISCUSSION

Overall, we observe significant site-related effects in multi-shell white matter tensor modeling from the shifting and scalar factors via ComBat-GAM. HBCD sets a strong example for harmonization steps during acquisition, though statistical methods such as ComBat-GAM further minimize scanner-related effects. We encourage future HBCD researchers to consider statistical methods to ensure maximum harmonization. Mitigation of these scanner-related effects will assist future efforts within and beyond the HBCD study to reliably map white matter trajectories throughout the lifespan [16], develop robust, pediatric-specific white matter quantification methods, and study deviations in cases of disease and disorder [11]. However, the work to mitigate scanner-induced site effects does not stop with this study. HBCD does not specify Δ (diffusion sensitization time) and δ (gradient pulse duration) within its scanning protocol, and hardware constraints introduce heterogeneity in these parameters. Further work to identify and mitigate variation within individual sites remains necessary to control for possible non-protocol variations, and future harmonization efforts from this cohort should include a longitudinal data as it becomes available.

## 5. COMPLIANCE WITH ETHICAL STANDARDS

This research study was conducted using human subject data made available by HBCD in accordance with all data use agreements therein.

## 6. ACKNOWLEDGMENTS

Data used in the preparation of this article were obtained from the HEALthy Brain and Child Development (HBCD) Study (https://hbcdstudy.org/), held in the NIH Brain Development Cohorts Data Sharing Platform. This is a multisite, longitudinal study designed to recruit approximately 7,000 families and follow them from pregnancy to early childhood. A full list of participating sites is available at: https://hbcdstudy.org/recruitment-sites/. HBCD Study Consortium investigators designed and implemented the study and/or provided data but did not necessarily participate in the analysis or writing of this report. This manuscript reflects the views of the authors and may not reflect the opinions or views of the NIH or the HBCD Study Consortium investigators. The HBCD Study is supported by the NIH and additional federal partners under award numbers U01DA055352, U01DA055353, U01DA055366, U01DA055365, U01DA055362, U01DA055342, U01DA055360, U01DA055350, U01DA055338, U01DA055355, U01DA055363, U01DA055349, U01DA055361, U01DA055316, U01DA055344, U01DA055322, U01DA055369,

U01DA055358, U01DA055371, U01DA055359, U01DA055354, U01DA055370, U01DA055347, U01DA055357, U01DA055367, U24DA055325, and U24DA055330. A full list of supporters is available at https://hbcdstudy.org/federal-partners/. The HBCD data repository grows and changes over time. The HBCD data used in this report came from DOI 10.82525. DOIs can be found at https://www.nbdc-datahub.org/hbcd-release-1-1.

The collaboration for this project was made possible by the VALIANT REACH program. This work has been funded in part by NIH 5U01DA055347-03, NIH 1R01EB017230, U24AG074855, the National Science Foundation Graduate Research Fellowship (NSF GRFP), and P50HD103537. This work was supported by the Alzheimer's Disease Sequencing Project Phenotype Harmonization Consortium (ADSP-PHC) that is funded by NIA (U24 AG074855, U01 AG068057 and R01 AG059716). The content is solely the responsibility of the authors and does not necessarily represent the official views of the NIH.